\documentclass{revtex4}
\usepackage{psfig}
\usepackage{subfigure}
\usepackage{graphicx}
\usepackage{doublespace}
\usepackage{amsmath}

\begin{document}

\renewcommand{\baselinestretch}{2}


\doublespace

\author{Francesco Sorrentino${}^{*\ddagger}$, Edward Ott${}^{*}$ }
\affiliation{${}^*$ Institute for Research in Electronics and Applied Physics, Department of Physics, and Department of Electrical and Computer Engineering, University of Maryland, College Park, Maryland 20742\\${}^\ddagger$ Universit{\`a} degli Studi di Napoli Parthenope, 80143 Napoli, Italy}

\begin{abstract}
In this paper we consider networks of dynamical systems that evolve in synchrony and investigate how dynamical information from the synchronization dynamics can be effectively used to learn the network topology, i.e., identify the time evolution of the couplings between the network nodes. To this aim, we present an adaptive strategy that, based on a potential that the network systems seek to minimize in order to maintain synchronization, can be successfully applied to identify the time evolution of the network from limited information. This strategy takes advantage of the properties of synchronism of chaos and of the presence of different communication delays over the network links. As a motivating example we consider  a network of sensors surveying an area, in which information regarding the time evolution of the network connections can be used, e.g., to detect changes taking place within the area. We propose two different setups for our strategy. In the first one, synchronization has to be achieved at each node (as well as the identification of the couplings over the network links), based solely on a single scalar signal representing a superposition of signals from the other nodes in the network. In the second one, we incorporate an additional node, termed the maestro, having the function of maintaining network synchronization. We will see that when such an arrangement is realized, it will become possible to effectively identify the time evolution of  networks that are much larger than would be possible in the absence of a maestro.
\end{abstract}

\title{Using synchronism of chaos for adaptive learning of network topology}
\maketitle

\section{Introduction}

In recent years, much effort has been devoted to the study of synchronization of networks of coupled dynamical systems \cite{WinBOOK,KuraBOOK,Er:Ko1,exploring,Str2,Pe:Ca,Ba:Pe02,Report,Ni:Mo,paradox,Mo:synchro2,Oh:Rho,Lee05,Hild1,Zh:Ku06,Wang,Wang:Chen02,ADAP1,ADAP2,ADAP3,ADAP4,ADAP5,DELAY1,DELAY2}. Important results have been obtained under the assumption that communication between the connected systems is immediate; i.e., the signal from system $i$ at time $t$ is received by systems $j$ at the same time $t$. Most of these studies have also focused on the case where the network topology is static, i.e., the coupling strengths between the connected systems is fixed and does not evolve in time. Yet in real situations networks of dynamical systems are often time-varying and affected by communication delays.
In this paper we will address situations  incorporating both the time varying structure common in typical real networks and the delays which inherently affect communication across the network. Our main emphasis will be on the situation where the time-dependent connection strength from system $j$ to another system $i$ in the network is unknown by system $i$, and we investigate how system $i$ can deduce these strengths by an adaptive strategy making use of the phenomenon of synchronization of chaos. As we shall show, diversity in the delay time associated with different links, as well as chaos of the exchanged signals, will play a crucial role in achieving this goal.



As a possible motivation, we will consider as our reference application that of a network of sensors that are dispersed over a ground and interact with each other through direct wireless communication. Namely, we will assume that the sensor network is employed to detect the evolution of the configuration within the surveyed area, e.g., due to motions of objects on the ground. In that case, a movement of an object across a communication pathway affects the strength of the signals exchanged by the nodes at its endpoints. Therefore, if we can succeed in identifying the time-evolution of the network couplings, this will provide information potentially enabling us to localize the position of the objects moving across the ground. We consider that each of the dynamical systems (`nodes') are identical chaotic systems broadcasting chaotic signals. As the network evolves, each node uses an adaptive strategy (which we specify) to maintain synchronization, and, in so doing, it determines the strengths of the couplings to it. 

Moreover, we will also show how our adaptive strategy could greatly benefit from the introduction of a special node, having the specific function of maintaining network synchronization. In what follows, we will refer to this particular node as the \emph{maestro}, in order to distinguish it from the remaining nodes of the network, which we will refer to as the \emph{orchestra}. For example, for the case of our reference application, we will make the assumption that this node is positioned in such a way that the strength of the signal sent by this node is not affected by the movements of the objects on the ground 
(e.g., it might be located atop a tower, hill or tall building). We will see that when such an arrangement is realized, it will become possible to effectively identify the time evolution of much larger networks than would be possible in the absence of a maestro.




The problem of identifying the structure of real complex networks, arises in all those situations where the exact network topological structure is unknown or uncertain, therefore it is of relevance in various fields of science and technologies. Examples are metabolic networks, biological neural networks, electric power grids, and so on  \cite{REF4,REF1,REF2}. Moreover, in many real world situations, there is often an intrinsic difficulty in extracting information on the network structure. Hence, the problem of learning the network topology,  represents a challenge in a number of real applications, that involve both natural and man-made systems \cite{REF4,REF1,REF2,REF3}.

In Sec. II we present our proposed adaptive strategy for maintaining synchronization of chaos and determining network topology. We do this both for the case with no maestro and for the case with a maestro. We also show that it is essential for our technique that signals be chaotic and experience time delays along the links of the network. In Sec. III we test our strategy through numerical experiments. We show that the presence of a maestro can greatly benefit  our strategy. Conclusions and further discussion are presented in Sec. IV.

\section{Adaptive strategy}

In a previous paper \cite{SOTT} we showed how a time-varying complex network can be synchronized by using a `potential' that each network node seeks to minimize. 
In this paper, we will extend this concept with the goal, of not only maintaining synchronization, but of also identifying the time-varying evolution of the network itself.
 In Refs. \cite{Identif1,Identif2} it was shown how to identify the static structure of a complex network of synchronizing systems by introducing a response (replica) network that can adjust its couplings to converge to those of the original one. The main focus of our paper will be on identifying the evolution of time-varying couplings of a complex network from limited information. We will show how this may be   achieved by solving a small number of auxiliary differential equation, provided that some general conditions on the time delays of the network links are met. We will consider two versions of our problem. In \emph{Version I} the maestro is absent, while in \emph{Version II} we have a maestro.

 In \emph{Version I}, the set of dynamical equations describing the dynamical state vector $x_i(t)$ of dimension $n$ at each node $i$ of the time evolving network is the following:
\begin{equation}
\dot x_i(t) =  F(x_i(t)) + \Gamma  \{ r_i(t) - \sum_{j=1, j \neq i}^{N} \bar{A}_{ij}(t) H (x_i(t-\tau_{ij})) \},  \quad i=1,2,...,N, \label{NP}
\end{equation}
where
\begin{equation}
r_i(t)= \sum_{j=1, j \neq i}^{N} A_{ij}(t) H( x_j(t-\tau_{ij})) \label{ri}
\end{equation}
is the \emph{incoming signal} at node $i$, which is a linear combination of the different scalar \emph{transmitted signals} $H(x_j(t))$, through the time varying coupling coefficients  $A_{ij}(t)$.  Here $\Gamma=[\Gamma_1,\Gamma_2,...,\Gamma_n]^T$ is a constant $n$-vector specifying the coupling of the scalar received signal $r_i$ to node $i$. (In our numerical experiments in Sec. III, we will take the components of $\Gamma$ to be all zero except for one component, say $\ell$, for which $\Gamma_\ell>0$.) 
We assume that the time delays $\tau_{ij}$, between each pair of connected nodes $i$ and $j$,  are known, {e.g., the sensors occupy fixed positions and thus the communication delays can be easily obtained from knowledge of the relative distances between the sensors.}
In contrast, we assume that the strengths of the network interactions, represented by the $A_{ij}$, are unknown and variable in time. Here, the $A_{ij}(t)$'s might represent the spreading, scattering, and attenuation of a wireless output signal broadcast by node $j$ to node $i$. In the case where the input to $i$ from $j$ is accomplished by propagation of an electromagnetic wave from $j$ to $i$, the link strength $A_{ij}$ will vary with time as a body moves through the propagation path from $j$ to $i$. Although the $A_{ij}(t)$ vary with time, we shall be most interested in the case where the time scale for this variation is slow compared to the natural time scale of the chaotic dynamics of an individual uncoupled system,
\begin{equation}
\dot{x}_i(t)=F(x_i(t)). \label{is}
\end{equation}
The quantity $\bar{A}_{ij}$, appearing in Eq. (\ref{NP}) represents an estimate made at node $i$ of the true coupling $A_{ij}(t)$. How this estimate can be made, as well as the accuracy of the estimate, will be the main focus of this paper. 

In \emph{Version II}, we consider an additional dynamical state vector, i.e., $x_M(t)$, associated with the maestro node, and the following set of dynamical equations:
\begin{equation}
\dot x_i(t) =  F(x_i(t)) + \Gamma  \{ r_i -[\frac{\alpha}{N-1} \sum_{j=1, j \neq i}^{N} \bar{A}_{ij} H (x_i(t-\tau_{ij})) + (1-\alpha) A_{iM}(t) H(x_i(t-\tau_{iM}))]\}  ,  \quad i=1,2,...,N, \label{NP2}
\end{equation}
where
\begin{equation}
r_i(t)= \frac{\alpha}{N-1} \sum_{j=1, j \neq i}^{N} A_{ij}(t) H( x_j(t-\tau_{ij})) + (1-\alpha) A_{iM}(t) H(x_M(t-\tau_{iM})), \label{ri2}
\end{equation}
and the maestro obeys the same dynamical evolution as an uncoupled node, i.e.,
\begin{equation}
\dot x_M(t)=F(x_M(t)). \label{maestro}
\end{equation}
Here $A_{iM}(t)$ is the strength of the coupling of the maestro to each node $i=1,...,N$; $\tau_{iM}$ is the communication delay between the maestro and node $i=1,...,N$; $0\leq \alpha \leq 1$ is a parameter that tunes the relative strength of the coupling exerted on the network nodes by the maestro, with respect to that of the other nodes in the network (the orchestra). When $\alpha=0$, each node is affected only by the forcing by the maestro, and therefore it can be seen as an isolated slave system connected to the master. 
On the other hand, when $\alpha=1$, each node will not feel the influence of the maestro and synchronization can only be realized from the mutual coupling with the other nodes in the network (the orchestra); thus the case $\alpha=1$ corresponds to our problem in Version I. Values of $\alpha$ between $0$ and $1$ (the case of most interest) correspond to intermediate situations where each node feels both the influence of the maestro, as well as that of the orchestra. 

{In what follows we assume that $A_{iM}(t)$ is known a priori at each node $i$. In Sec. III.D we also address the case that the assumed known coupling with the maestro is affected by a small error.
}

In the particular case where we actually have the correct estimates $\bar{A}_{ij}=A_{ij}$, the network equations in (\ref{NP}) and (\ref{NP2}), can be rewritten as
\begin{subequations}
\begin{align}
  \dot x_i(t) =  F(x_i(t)) + \Gamma  \sum_{j=1, j \neq i}^{N} {A}_{ij}(t) \{ H(x_j(t-\tau_{ij}))- H (x_i(t-\tau_{ij})) \}, \qquad i=1,2,...,N, \qquad \mbox{for Version 1},\label{N1} \\
  \mbox{and    ~~~~~~~~~~~~~~~~~~~~~~~~~~~~~~~~~~~~~~~~~~~~~~~~~~~~~~~~~~~~~~~~~~~~~~~~~~~~~~~~~~~~~~~~~~~~~~~~~~~~~~~~~~~~~~~~~~~~~~~~~~~~~~~~~~~~~~                                                                           \nonumber} \\
  \dot x_i(t) =  F(x_i(t)) + \Gamma \{ \frac{\alpha}{N-1} \sum_{j=1, j \neq i}^{N} {A}_{ij}(t) \{ H(x_j(t-\tau_{ij}))- H (x_i(t-\tau_{ij})) \} + \nonumber \\ + (1-\alpha) A_{iM}(t) \{H(x_M(t-\tau_{iM}))-H(x_i(t-\tau_{iM}))\} \} ,\label{N2}  \qquad i=1,2,...,N, \qquad \mbox{for Version 2}.
\end{align}
\end{subequations}
Thus, in a similar way to the case of a network of coupled oscillators without delays or time variability of the coupling $A_{ij}$ \cite{Pe:Ca}, the synchronization manifold,
\begin{subequations}
\begin{align}
x_1=x_2=...=x_N, \label{SM} \qquad \mbox{for Version 1}, \\
x_1=x_2=...=x_N=x_M, \label{SM2} \qquad \mbox{for Version 2},
\end{align}
\end{subequations}
represents an invariant set for the set of equations (\ref{N1}) or (\ref{N2}). 

Once the dynamical functions $F$ and $H$ are assigned,  the stability of synchronized states to perturbations from the synchronization manifold will depend on the dynamical evolution of the network adjacency matrix $A(t)$, on the overall coupling $\Gamma$, and on the  time delays $\tau_{ij}$ (plus on $A_{iM}(t)$ and $\tau_{iM}$ in the case of Version 2).  In what follows, we will proceed under the assumption that the synchronization manifold (\ref{SM} - \ref{SM2}) is stable for our choices of the time evolution of $A(t)$, $A_{iM}(t)$, $\Gamma$ and the time delays $\tau_{ij}$ and $\tau_{iM}$, and we will focus on the problem of identifying the network structure (i.e., obtaining the estimates $\bar{A}_{ij}(t)$).

We assume that at each node $i$ direct information on the couplings $A_{ij}(t)$ is unavailable. Thus, the estimates $\bar{A}_{ij}$ at node $i$ must be obtained solely using the aggregate incoming signal $r_i(t)$. In order to make this estimate, we introduce a `potential' $\Psi_i$ at each node $i$:
\begin{subequations}
\begin{align}
\Psi_i(t)=  \nu \int^t e^{-\nu (t-t')} \{ r_i(t') - \sum_{j \neq i} \bar{A}_{ij}(t) H( x_i(t'-\tau_{ij})) \}^2 dt', \label{tri_or}\qquad \mbox{for Version 1}, \\
\Psi_i(t)=  \nu \int^t e^{-\nu (t-t')} \{ r_i(t') - \frac{\alpha}{N-1} \sum_{j \neq i} \bar{A}_{ij}(t) H( x_i(t'-\tau_{ij})) -(1-\alpha) A_{iM}(t) H(x_i(t'-\tau_{iM})\}^2 dt', \label{tri_or2}\qquad \mbox{for Version 2}.
\end{align}
\end{subequations}
If the time scale for variation of $\bar{A}_{ij}$ is larger than $\nu^{-1}$, then $\Psi_i(t)$ can be approximated by
\begin{subequations}
\begin{align}
\Psi_i(t) \cong  \nu \int^t e^{-\nu (t-t')} \{ r_i(t') - \sum_{j \neq i} \bar{A}_{ij}(t') H( x_i(t'-\tau_{ij})) \}^2 dt', \label{tri}\qquad \mbox{for Version 1}, \\
\Psi_i(t) \cong  \nu \int^t e^{-\nu (t-t')} \{ r_i(t') - \frac{\alpha}{N-1} \sum_{j \neq i} \bar{A}_{ij}(t') H( x_i(t'-\tau_{ij})) -(1-\alpha) A_{iM}(t') H(x_i(t'-\tau_{iM})\}^2 dt', \label{tri2} \qquad \mbox{for Version 2}.
\end{align}
\end{subequations}
As given by (\ref{tri}) and (\ref{tri2}), the potential $\Psi_i$ is approximately the time averaged squared synchronization error, and $\nu^{-1}$ is the temporal extent over which the averaging is performed. Note from (\ref{tri}) that $\Psi_i(t) \cong 0$ if $ \sum_{j \neq i} \bar{A}_{ij}(t) H( x_i(t-\tau_{ij})) = \sum_{j \neq i} {A}_{ij}(t) H( x_j(t-\tau_{ij})) $. Similarly, from (\ref{tri2}) we note that $\Psi_i(t) \cong 0$ if $ \sum_{j \neq i} \bar{A}_{ij}(t) H( x_i(t-\tau_{ij})) = \sum_{j \neq i} {A}_{ij}(t) H( x_j(t-\tau_{ij})) $ and $x_i(t-\tau_{iM})=x_M(t-\tau_{iM})$. Due to the chaotic nature of the $x$'s, provided that the $\tau_{ij}$ are all distinct, these conditions can only be satisfied if $\bar{A}_{ij}(t)={A}_{ij}(t)$ (i.e., our estimates of the time evolutions of the network couplings are correct) and $x_i(t-\tau_{ij})=x_j(t-\tau_{ij})$ (i.e., the network systems are synchronized in time). 
Thus we seek to minimize $\Psi_i$. Therefore, we impose that the $\bar{A}_{ij}(t)$ evolve according to the following gradient descent relation:

\begin{subequations}
\begin{align}
\frac{d \bar{A}_{ij}}{dt}= - \beta \frac{d \Psi_{i}}{d \bar{A}_{ij}}= 2 \beta ({c_{ij} - \sum_{k \neq i} \bar{A}_{ik} C_{ijk}}), \label{sei}\qquad \mbox{for Version 1}, \\
\frac{d \bar{A}_{ij}}{dt}= - \beta \frac{d \Psi_{i}}{d \bar{A}_{ij}}= 2 \frac{\beta \alpha}{N-1} [ {c_{ij}+ (1-\alpha)\xi_{ij} - \frac{\alpha}{N-1}\sum_{k \neq i} \bar{A}_{ik} C_{ijk}} ], \label{sei2} \qquad \mbox{for Version 2},
\end{align}
\end{subequations}
where
\begin{equation}
c_{ij}(t)= \nu \int_0^{t} e^{- \nu (t-t')} r_i(t') H(x_i(t'-\tau_{ij})) dt', \label{cpi}
\end{equation}
\begin{equation}
C_{ijk}(t)= \nu \int_0^{t} e^{- \nu (t-t')} H(x_i(t'-\tau_{ij})) H(x_i(t'-\tau_{ik})) dt', \label{cgr}
\end{equation}
\begin{equation}
\xi_{ij}(t)= \nu  \int_0^{t} e^{- \nu (t-t')}  A_{iM}(t') H(x_i(t'-\tau_{ij})) H(x_i(t'-\tau_{iM})) dt', \label{cs}
\end{equation}
and $\beta$ is a constant larger than zero.

In practice it is useful to avoid doing the above integrations at each time step of our adaptive computation. Rather, we note that $c_{ij}(t)$, $C_{ijk}(t)$, and $\xi_{ij}(t)$ given by (\ref{cpi}) and (\ref{cgr}), and (\ref{cs}) satisfy simple differential equations, and we employ these in our adaptive computation of $\bar{A}_{ij}(t)$:
\begin{equation}
\frac{ d c_{ij}(t)}{dt}= - \nu c_{ij}(t) + \nu r_i(t) H( x_i(t-\tau_{ij}) ), \label{sette}
\end{equation}
\begin{equation}
\frac{ d C_{ijk}}{dt}= - \nu C_{ijk}(t) + \nu H( x_i(t-\tau_{ij}) ) H( x_i(t-\tau_{ik}) ), \label{otto}
\end{equation}
\begin{equation}
\frac{ d \xi_{ij}}{dt}= - \nu \xi_{ij}(t) + \nu  A_{iM}(t) H( x_i(t-\tau_{ij}) ) H( x_i(t-\tau_{iM}) ) \label{nove}.
\end{equation}


The simplest version of this setup, used in our numerical example in Sec. III, corresponds to letting $\beta \rightarrow \infty$, in which case Eqs. (11) implies that we can obtain the $\bar{A}_{ik}$ by solving the system of linear equations,
\begin{subequations}
\begin{align}
c_{ij}=\sum_{k \neq i} \bar{A}_{ik} C_{ijk}, \qquad \mbox{for Version 1}, \\
c_{ij}-(1-\alpha)\xi_{ij}=\frac{\alpha}{N-1}\sum_{k \neq i} \bar{A}_{ik} C_{ijk} \qquad \mbox{for Version 2},
\end{align}
\end{subequations}
where $c_{ij}$, $C_{ijk}$, and $\xi_{ij}$ are obtained by solving (\ref{sette}), (\ref{otto}), and (\ref{nove}).


According to our previous discussion (in particular, the condition for the approximate equality in Eq. (\ref{tri})), in order for our identification strategy to work properly we should choose $\nu$ such that it is in the intermediate range,
\begin{equation}
T_s < \nu^{-1}< T_{n}, \label{inu}
\end{equation}
where $T_s$ is the time scale of the chaotic dynamics of an uncoupled nodal system ($\dot{x}=F(x)$), and $T_n$ is the time scale on which the network itself changes (i.e. the time scale on which $A_{ij}(t)$ varies). We will subsequently give quantitative definitions of $T_s$ and $T_n$ for our examples.

We emphasize that we try to identify the time evolutions of the \emph{several} $A_{ij}$'s, by using only \emph{one} piece of external information at each node $i$, namely, the incoming signal $r_i(t)$. Our ability to extract estimates for the $N-1$ values of $A_{ij}(t)$ ($i=1,2,...,N; j \neq i$) at each node $i$ from each incoming signal, relies on the temporal signal diversity of $r_i(t)$ due to the presence of both chaos and \emph{different} time delays along the network links. To illustrate the necessity of different time delays, consider the following example. Assume, for instance, that our network consists of three coupled systems in the Version I scenario ($N=3$ in Eq. (1)), and we want to identify the time evolutions of $A_{12}(t)$ and $A_{13}(t)$ when the only available information at node 1 is $r_1(t)$, the time evolution of the incoming signal at node 1. If we are on the synchronization manifold, then Eqs. (1) and (2) yield
\begin{equation}
\dot{x}_1(t)=F(x_1(t))+ \Gamma \{ [A_{12}(t)-\bar{A}_{12}(t)] H(x_1(t-\tau_{12}))+ [A_{13}(t)-\bar{A}_{13}(t)] H(x_1(t-\tau_{13})) \},
\label{unoduetre} \end{equation}
and, for $\tau_{12}=\tau_{13}$, the $H$ terms are equal, yielding for the term in (\ref{unoduetre}) that is in curly brackets
\begin{equation}
 H(x_1(t-\tau_{12})) \{[A_{12}(t)-\bar{A}_{12}(t)] - [A_{13}(t)-\bar{A}_{13}(t)]  \}.
\end{equation}
Thus, if $\tau_{12} = \tau_{13}$, we see that we could obtain synchronization at node $1$ for any $\bar{A}_{12}(t)$ and $\bar{A}_{13}(t)$ such that $\bar{A}_{12}(t)+\bar{A}_{13}(t)={A}_{12}(t)+{A}_{13}(t)$, and we cannot hope to estimate \emph{both} ${A}_{12}(t)$ and ${A}_{13}(t)$ only from $r_1(t)$. In contrast, if $\tau_{12} \neq \tau_{13}$, then on the synchronization manifold $x_2(t-\tau_{12}) \neq x_3(t-\tau_{13})$ and the previous degeneracy is removed. Thus we now have the possibility that it may be feasible to obtain the desired two unknowns $A_{12}(t)$ and $A_{13}(t)$;  in other words,  it may become possible to obtain good estimates $\bar{A}_{12}(t)$ and $\bar{A}_{13}(t)$, provided that $\tau_{12} \neq \tau_{13}$. Hence the presence of different delays is crucial in the identification process.

Moreover our strategy benefits from the chaotic nature of the exchanged signals at the network nodes. In particular, we take advantage of the high information content encoded in chaotic signals. As an illustrative counterexample, assume that, in place of the chaotic signals, we had sine waves of the same frequency broadcast by the network systems. The communication delays would result in phase shifts of the sine waves, and the signal received at each node, being the sum of sinusoids of the same frequency, would be a sinusoidal signal. In this case, there would be no information content in the received  signal other than its amplitude and phase; therefore, we would not be able to extract more than two pieces of information from it.

\section{Numerical experiments}

 \subsection{Version I}
As a first example, we consider the problem in Version I [Eqs. (\ref{NP}) and (\ref{ri})], and we seek to estimate the time evolution of a small network of $N=3$ nodes and $M=6$ directed links (two directed links incoming to each node).
For our uncoupled dynamical system, $\dot{x}_i=F(x_i)$, we consider a R\"ossler oscillator, $x_i=(x_{i1},x_{i2},x_{i3})^T$ and
\begin{equation}
F(x_i)=\left[\begin{array}{c}
    -x_{i2}-x_{i3} \\
    x_{i1} + 0.165 x_{i2} \\
    0.2+ (x_{i1}-10 )  x_{i3}
  \end{array} \right]. \label{F}
\end{equation}
We choose the oscillators to be coupled in the $x_{i1}$ variable, i.e., $H(x)=x_1$, $\Gamma=[\Gamma_1,0,0]^T$, $\Gamma_1>0$.

The delays associated with links $j \rightarrow i$, are given in  matrix form, $\mathcal{T}=\{ \tau_{ij} \}$, as follows:
\begin{equation}
\mathcal{T}=0.3 T_{s} \left[\begin{array}{cccc}
    0 &  1 & 2   \\
    1 & 0 & 3    \\
    2 &  3 & 0 \\
  \end{array} \right], \label{T2}
\end{equation}
where $T_s$ is defined as the time $s$ at which the autocorrelation function $\mathcal{C}(s)$ of $x_{s1}(t)$, $\mathcal{C}(s)=<x_{s1}(t+s) x_{s1}(t)>-<x_{s1}(t)>^2$, decays to $\mathcal{C}(0)/2$, and $x_s(t)$ is a typical orbit on the chaotic attractor of the uncoupled system, $\dot{x}_s=F(x_s)$. Note that for each $i$, the  $\tau_{ij}$ for every $j \neq i$ are all unequal. For our choice in Eq. (\ref{F}), we obtain $T_s=0.75$.

%


For $t>0$ we assume the following network evolution,
\begin{equation}
A_{ij}(t)=1+ \epsilon_{ij} \sin (\omega_{ij}t+ \phi_{ij}^0),  \label{net_ev}
\end{equation}
where the $\epsilon_{ij}=\epsilon_{ji}$ are random numbers drawn from a uniform distribution between $0$ and $1$, the $\omega_{ij}=\omega_{ji}$ are random numbers drawn from a uniform distribution between $\omega_{min}$ and $\omega_{max}$, and the $\phi_{ij}^0=\phi_{ji}^0$ are random numbers drawn from a uniform distribution between $0$ and $2 \pi$. We define the network evolution time scale $T_n$ for our example to be $T_n=\omega_{max}^{-1}$.

The network state vectors are initialized as follows,
\begin{equation}
x_{i1}^0= x_1^0 + c \rho_1 \epsilon_{ix}, \quad  x_{i2}^0= x_2^0 + c \rho_2 \epsilon_{iy}, \quad  x_{i3}^0= x_3^0 + c \rho_3 |\epsilon_{iz}|. \label{IC}
\end{equation}
 Here $(x_1^0,x_2^0,x_3^0)$ is a randomly chosen point on the R\"ossler attractor, obtained by evolving the uncoupled system, $\dot {x}_s=F(x_s)$, until it is on the attractor, and then taking its state at some arbitrarily chosen subsequent random time;  $\epsilon_{ix}, \epsilon_{iy}$ and $\epsilon_{iz}$ are zero-mean independent random numbers of unit variance drawn from a normal distribution; $\rho_1=7.45,\rho_2=7.08,\rho_3=4.25$ are the standard deviations of the time evolutions of $x_{s1}(t),x_{s2}(t),x_{s3}(t)$ obtained from numerical solution of an uncoupled oscillator, $\dot {x}_s=F(x_s)$. 
 For our numerical experiment we take $\Gamma_1=0.25$, $\nu={1 \over{8 T_s}} \simeq 0.17$, $\omega_{max}=0.4 \times 10^{-4}$ and $\omega_{min}=\omega_{max}/2$. Using these parameters we have that the inequalities in Eq. (\ref{inu}) are satisfied: $0.75<6< 2.5 \times 10^4$. The computations were done over a time interval $0 \leq t \leq 10^5$.

To measure the extent to which our strategy works we introduce the following two error measures:
\begin{equation}
E_{x_1}(t)=\frac{1}{N \rho_1} { \sum_i{ |x_{i1}(t)-<{x}_{i1}(t)>|} },
\end{equation}
{where $<{x}_{i1}(t)>=\frac{1}{N} { \sum_i{ x_{i1}(t)}}$ and,}
\begin{equation}
E_{A}(t)=\frac{1}{M} { \sum_{i} \sum_{j \neq i} { |A_{ij}(t)-\bar{A}_{ij}(t)|} }.
\end{equation}

\begin{figure}[h]
\centerline{\psfig{figure=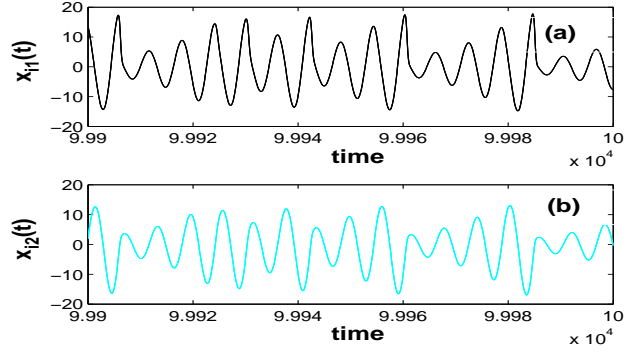,width=9cm,height=5cm}}
\caption{\small Time evolutions of ${x_{i1}(t)},{x_{i2}(t)}$ between $t=99,900$ and $t=100,000$. The plots for $i=1,2,3$ are identical to within the width of the plotted lines. \label{A1}}
\end{figure}

\begin{figure}[h]
\centerline{\psfig{figure=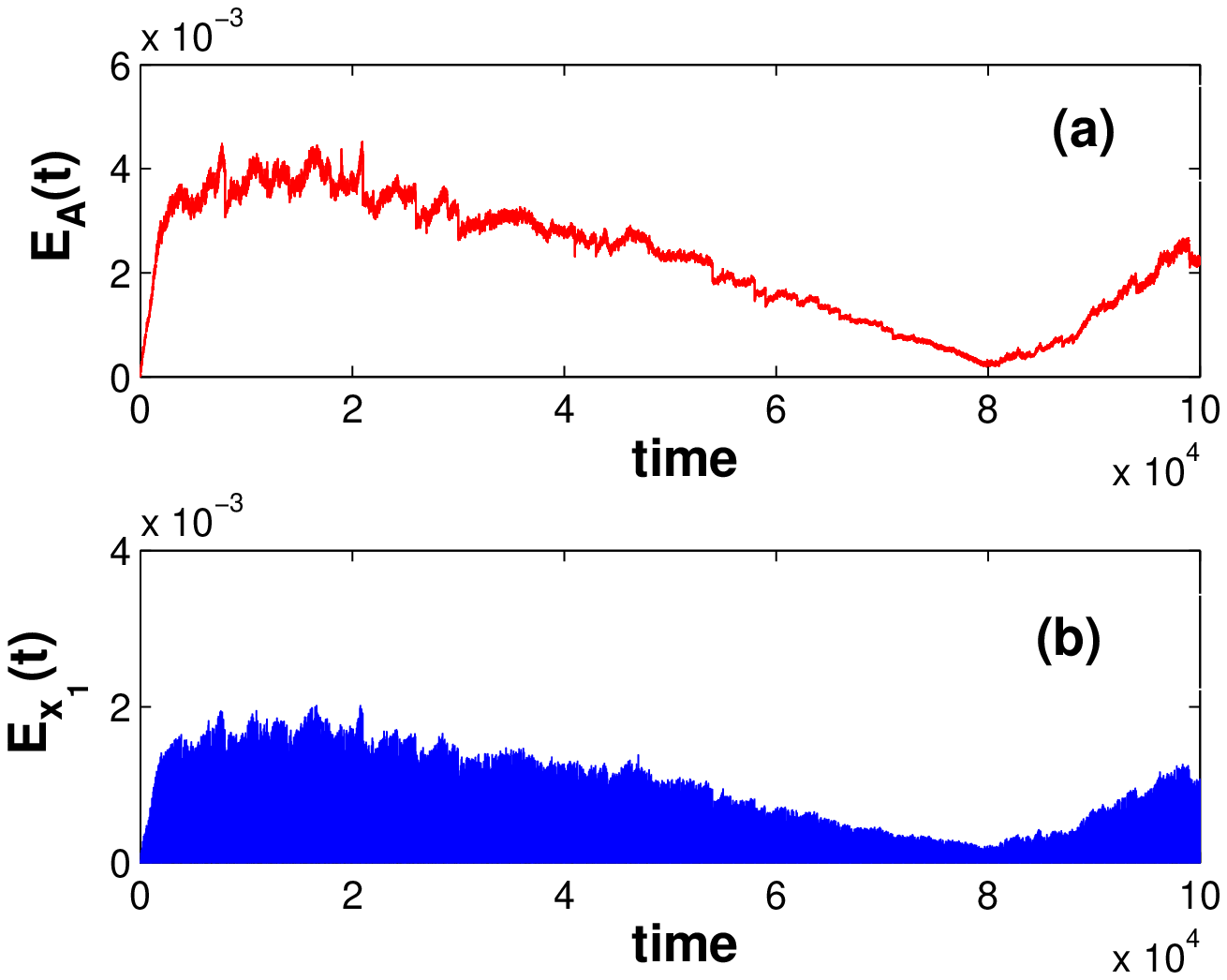,width=9cm,height=5cm}}
\caption{\small Time evolutions of $E_A(t)$ and $E_{x_1}(t)$, for $t$ between $0$ and $10^5$. 
\label{A2}}
\end{figure}

\begin{figure}[h]
\centerline{\psfig{figure=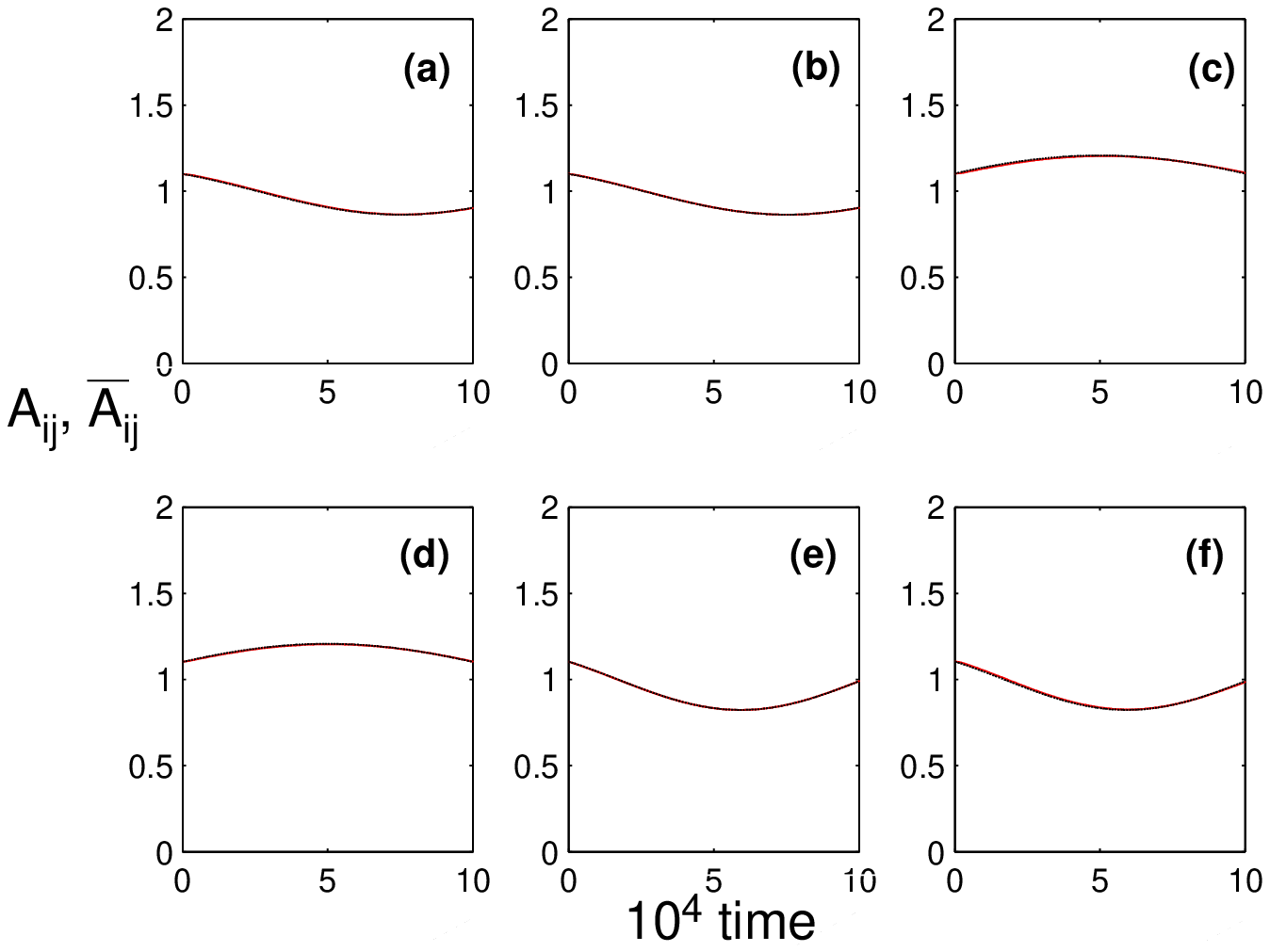,width=12cm,height=8cm}}
\caption{\small (Color online)  The time evolutions of $A_{ij}(t)$  and $\bar{A}_{ij}(t)$ are shown from $t=0$ to $t= 10^5$ and agree to within the widths of the plotted lines. Each subplot represents a directed link in the network: (a) i=1, j=2; (b) i=2, j=1; (c) i=3; j=1; (d) i=1, j=3; (e) i=2, j=3; (f) i=3 j=2. $\nu={1 \over{8 T_s}} \simeq 0.17$,  $\omega_{max}=0.4 \times 10^{-4}$ and $\omega_{min}=\omega_{max}/2$.
\label{A3}}
\end{figure}

Figures \ref{A1},\ref{A2}, and \ref{A3} show results of our computations. Figure \ref{A1} shows $x_{i1}(t)$ (Fig. \ref{A1}(a)) and $x_{i2}(t)$ (Fig. \ref{A1}(b)) versus $t$ at the end of the the run, $0.999 \times 10^5 \leq t \leq 10^5$. For all nodes, $i=1,2,3$, $x_{1i}(t)$ and $x_{2i}(t)$ are identical to within the width of the plotted curves. Figure \ref{A2}(a) shows that $E_A(t)$ is typically less than $0.5 \%$ over the entire run, while Fig. \ref{A2}(b) shows that the synchronization error $E_{x_1}(t)$ is typically less than $0.2 \%$. These results confirm that our adaptive strategy is effective in synchronizing the network and that the $\bar{A}_{ij}(t)$ closely follow the evolutions of the time evolution of the true couplings ${A}_{ij}(t)$ (see Fig. \ref{A3}).

\begin{figure}[h]
\centerline{\psfig{figure=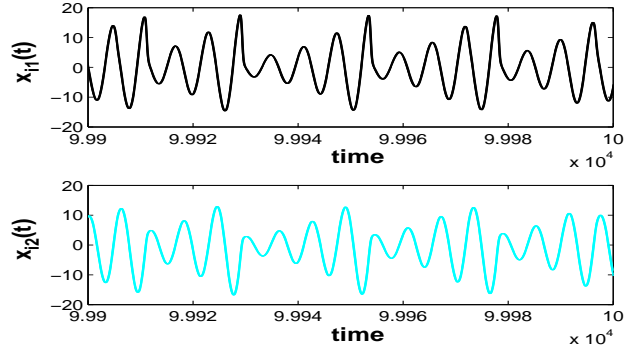,width=9cm,height=5cm}}
\caption{\small Time evolutions of ${x_{i1}(t)},{x_{i2}(t)}$ between $t=99,900$ and $t=100,000$. The plots for $i=1,2,3,4,5$ show almost-synchronized behavior. 
\label{F1}}
\end{figure}

\begin{figure}[h]
\centerline{\psfig{figure=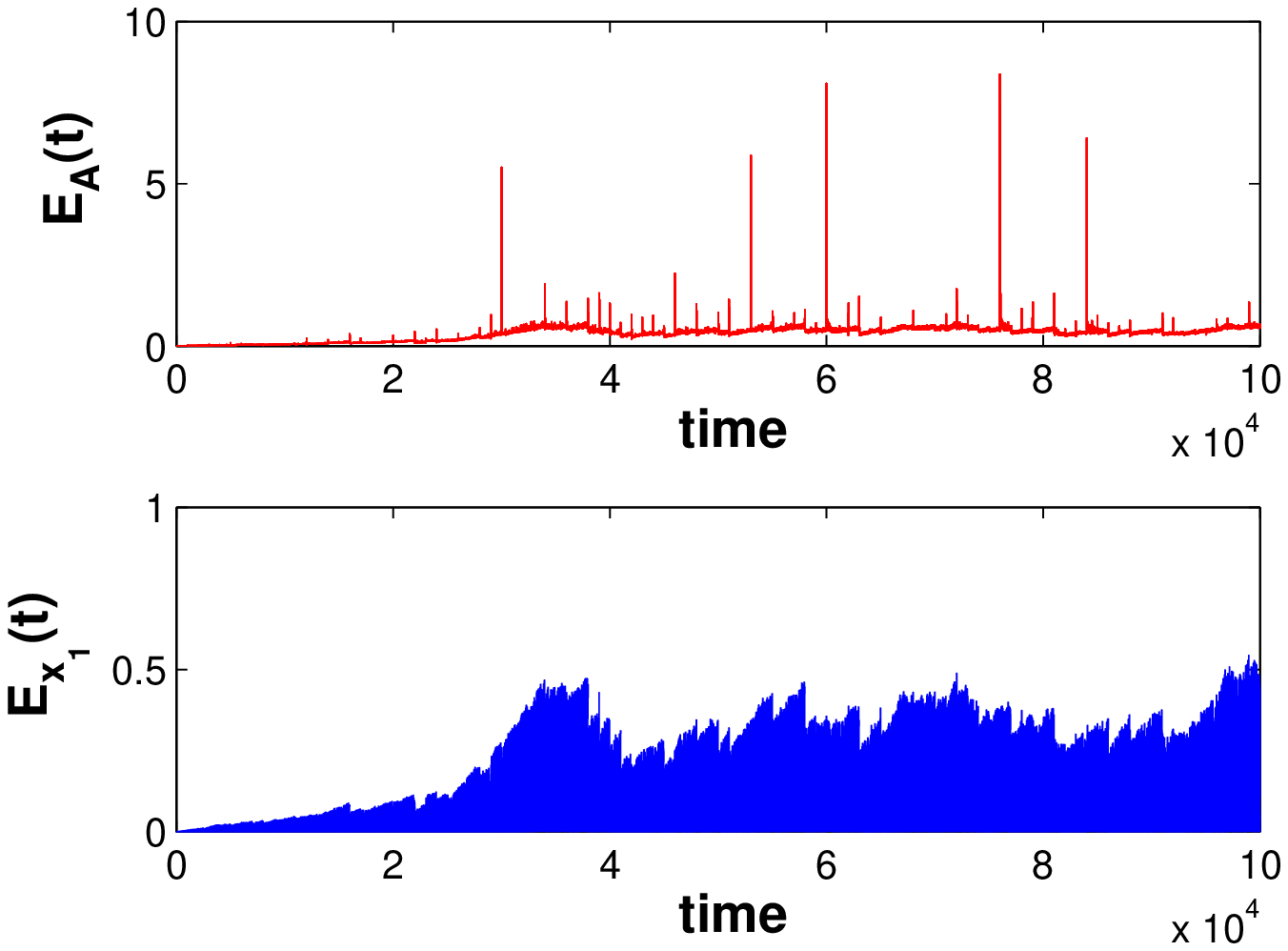,width=9cm,height=5cm}}
\caption{\small Time evolutions of $E_A(t)$ and $E_{x_1}(t)$, for $t$ between $0$ and $10^5$. 
\label{F2}}
\end{figure}

\begin{figure}[h]
\centerline{\psfig{figure=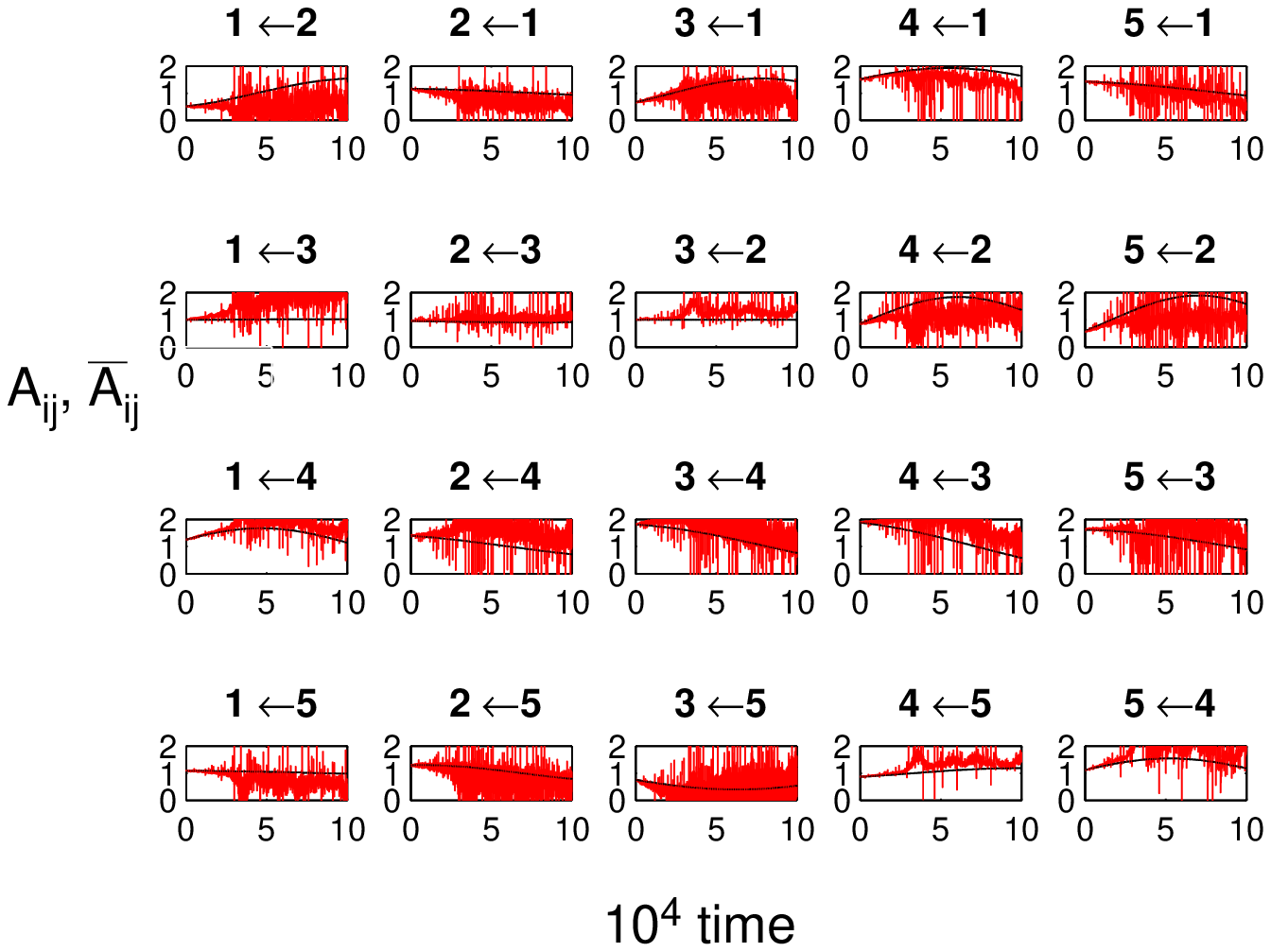,width=12cm,height=8cm}}
\caption{\small (Color) Each subplot represents a directed link in the network. The time evolutions of $A_{ij}(t)$ (black) and $\bar{A}_{ij}(t)$ (red) are plotted from $t=0$ to $t= 10^5$, with $\nu={1 \over{8 T_s}} \simeq 0.17$,  $\omega_{max}=0.4 \times 10^{-4}$ and $\omega_{min}=\omega_{max}/2$.
\label{F3}}
\end{figure}

Figures  \ref{F1}, \ref{F2}, and \ref{F3} show the results of another simulation where we considered a network of $N=5$ nodes and
\begin{equation}
\mathcal{T}=0.3 T_{s} \left[\begin{array}{ccccc}
    0 &  4 & 5 & 1 & 2\\
    4 & 0 & 1  & 2 & 3\\
    5 &  1 & 0 & 3 & 4\\
    1 & 2 & 3 & 0  & 5\\
    2 & 3 & 4 & 5  & 0\\
  \end{array} \right], \label{T2}
\end{equation}
with all the other conditions the same as those used in obtaining Figs. 1-3. As can be seen from Fig. \ref{F2}, for this case our identification strategy fails, with the $\bar{A}_{ij}$ deviating from the evolution of the ${A}_{ij}$. At the same time, we observe from Fig. \ref{F1} that the network systems evolve approximately synchronously in time. Thus, in this case we can approximately synchronize the network, but we fail in correctly estimating the evolution of the network couplings.

Unfortunately, by using the strategy in Version I, we were not able to synchronize complete graphs of R\"ossler oscillators, as the network dimension $N$ was increased. 
As we will explain in the Appendix of this paper, we have an understanding of the reason for this failure. Briefly, a requirement for correctly estimating $A_{ij}$ at any given node, say $i$, is that the $\tau_{ij}$'s associated with $i$ are different enough  (we address this issue more rigorously in the Appendix). 
However, if $N$ is too large, we cannot make the $\tau_{ij}$ too separated, 
because stability of the synchronization manifold (\ref{SM}) (under the condition $\bar{A}_{ij}=A_{ij}$), is lost if the $\tau_{ij}$'s are too large. Thus, for our setup in Version I, achieving the required diversity in the $\tau_{ij}$'s, contrasts with our requirement of having the network synchronize.

We have also tested our Version I scenario for a case as in Figs. 1-3, but with the network evolving on a faster time scale $\omega_{max}=2 \times 10^{-4}$ (instead of $0.4 \times 10^{-4}$), and for this faster evolution we have found that our strategy fails (results not shown).

\subsection{Version II}

The failure of our method with large $N$ when applied to Version I is the main reason for considering in what follows the alternative set up of our problem in Version II [Eqs. (\ref{NP2}) and (\ref{ri2})], which incorporates a maestro node, having the function of maintaining the network synchronizability.  Thus we assume that each node/dynamical system is subject to two actions: a \emph{coupling exerted by the other nodes of the network}, which in what follows we will refer to as \emph{the orchestra} and a \emph{coupling exerted by a master system}, which in what follows we will refer to as \emph{the maestro}. By tuning the parameter $\alpha$ in (\ref{NP2}) and (\ref{ri2}) between $0$ and $1$ we are able to interpolate between a situation where the nodes feel only the influence of the maestro and a situation where they feel only the influence of the orchestra (corresponding to Version I).

We assume that the maestro coupling $A_{iM}$ is known and time independent, but that the other network couplings, $A_{ij}$, are time-varying and unknown at the network nodes. In order to guarantee that the synchronization manifold is stable, we take the time delays with the maestro $\tau_{iM}$ to be small enough. Our hope is that, once synchronization is ensured through the maestro, we will potentially be able to choose the other $\tau_{ij}$'s so that they are sufficiently spread that we have a much greater likelihood of being able to correctly estimate the $A_{ij}(t)$. As we will show, by making use of the maestro node, we will be able to identify larger networks, whose time-evolution is faster than would be possible in Version I ($\omega_{max}> 10^{-4}$). 


To test our strategy in Version II, Eq. (\ref{NP2}), again, as for Version I, we assume that we want to identify the time evolution of globally connected networks of  R\"ossler oscillators, with $H(x)=x_1$, $\Gamma=[\Gamma_1,0,0]$, $\Gamma_1>0$. For our first experiment we take a network of $N=5$ nodes and $M=20$ directed links to be identified. We assume  $\tau_{iM}=0.1$, $A_{iM}(t)=1$ at each nodes $i$ in (\ref{NP2}); we have verified beforehand that in the case where $A_{ij}=\bar{A}_{ij}$ for $i,j=1,...,N$, for this choice of the parameters, the synchronization manifold (\ref{SM2}) is stable. We are interested in identifying the time evolutions of all the $A_{ij}$'s associated with the network links. 

We take the time delays $\tau_{ij}$ along the network directed links $i \rightarrow j$ to be
\begin{equation}
\tau_{ij}=\tau_{iM}+[1+mod(i+j,N)] \theta T_s,   \qquad i \neq j \label{ftau}
\end{equation}
 where $mod(a,b)$ is the remainder of the integer division of $a$ by $b$, and $\theta$ is a variable parameter to be specified. Note that, for our choice of the time delays in (\ref{ftau}), we have that $\tau_{ij}=\tau_{ji}$, $i \neq j$.
Moreover, from (\ref{ftau}) we also have that for any node $i$ and for any pair of delays associated with any two links incoming to $i$, say $\tau_{ij}$ and $\tau_{ik}$, $j \neq k$ we have that $\tau_{ij} \neq \tau_{ik}$, and the values $\tau_{ij}$ and $\tau_{ik}$ are separated by an integer multiple of $\theta T_s$. Thus, we do not have equal delays 
along incoming links to any node.

\begin{figure}[h]
\centerline{\psfig{figure=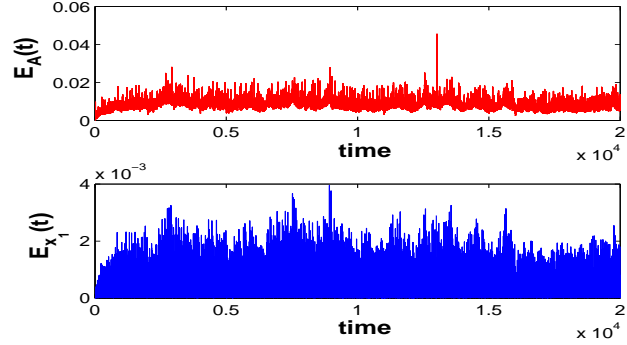,width=9cm,height=5cm}}
\caption{\small Time evolutions of $E_A(t)$ and $E_{x_1}(t)$, for $t$ between $t=0$ and $t=2 \times 10^4$, $\alpha=0.2$, $\nu={{8 T_s}^{-1}} \simeq 0.17$,  $\theta=0.6$, $\Gamma_1=1$, $\omega_{max}=4 \times 10^{-3}$, $\omega_{min}=\omega_{max}/2$.  \label{NPA2}}
\end{figure}

\begin{figure}[h]
\centerline{\psfig{figure=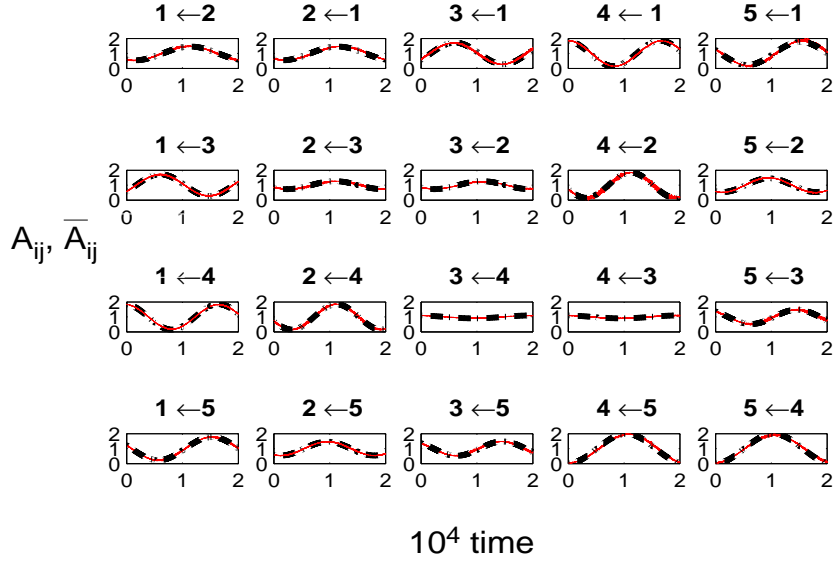,width=12cm,height=8cm}}
\caption{\small (Color) Each subplot represents a directed link in the network. The time evolutions of $A_{ij}(t)$ (in black dots) and $\bar{A}_{ij}(t)$ (in red curves) are plotted from $t=0$ to $t=2 \times 10^4$.   \label{NPA3}}
\end{figure}

 A first experiment involving our strategy in Version II, is shown in Figures \ref{NPA2} and \ref{NPA3}. For this experiment, we took $\alpha=0.2$  (i.e., synchronization depends mostly on the coupling with the maestro), and set $\theta=0.6$, indicating that the separation time between any pair of network delays is a multiple of $0.45=0.6 T_s$. 
Moreover we assumed  that the network evolves much faster than in the cases reported in Figs. 1-6 of Sec. III.A for Version I, that is, $\omega_{max}=4 \times 10^{-3}$, and $\omega_{min}=\omega_{max}/2$. The network was evolved  for a long time until time $2 \times 10^4$, starting from synchronous initial conditions and assuming $\bar{A}_{ij}(0)=A_{ij}(0)$.

\begin{figure}[h]
\centerline{\psfig{figure=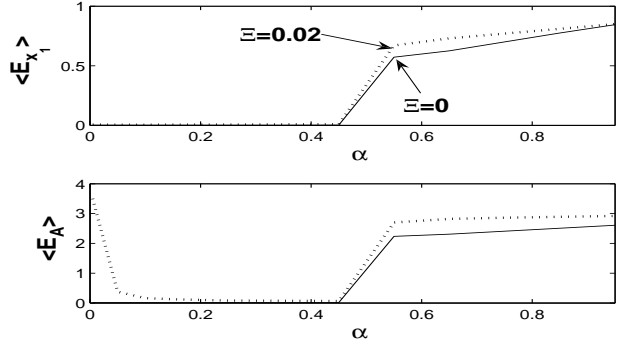,width=9cm,height=5cm}}
\caption{\small (Continuous line) The time averaged errors $<{E}_{x_1}>$ and $<{E}_{A}>$ are plotted versus $\alpha$, for  $N=5$, $\Gamma_1=1$, $\theta=0.6$, $\nu=1/(8 T_s)$, $\omega_{max}=4 \times 10^{-4}$, $\omega_{min}=\omega_{max}/2$. (Dashed line) The same experiment has been repeated for a case where a small error in the coupling with the maestro is present, i.e., $\Xi=0.02$ (see Sec. III.D). \label{alpha}}
\end{figure}

As shown in Fig. \ref{NPA2}, $E_A(t) <0.05$ and $E_{x_1}(t) < 4 \times 10^{-3}$, throughout the length of the run. The black dots in Fig. \ref{NPA3} show the time evolutions of the $A_{ij}$'s and the red lines those of the $\bar{A}_{ij}$'s. As can be seen, our adaptive strategy is effective in synchronizing the network with the $\bar{A}_{ij}(t)$ closely following the evolutions of the ${A}_{ij}(t)$. Remarkably, we have succeeded in identifying the time evolutions of \emph{all the $M=20$} links present in the network.
The same experiment in Figs. \ref{NPA2} and \ref{NPA3} has also been repeated for different values of $\alpha$. In Fig. {\ref{alpha}}, the time averaged errors $<{E}_{x_1}>$ and $<{E}_{A}>$ are plotted as $\alpha$ is increased between $0.005$ and $0.95$ (continuous line). As can be seen, our strategy is observed to be effective in identifying the network evolution for values of $\alpha$ up to about $0.45$.

\begin{figure}[h]
\centerline{\psfig{figure=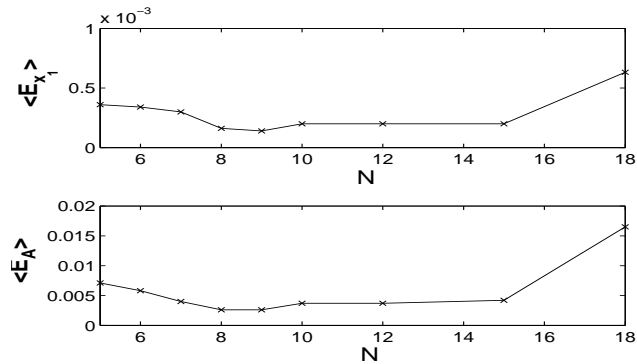,width=9cm,height=5cm}}
\caption{\small The time averaged errors $<{E}_{x_1}>$ and $<{E}_{A}>$ versus the network size $N$ for $5 \leq N \leq 18$, with  $\alpha=0.2$, $\Gamma_1=1$, $\theta=0.6$, $\nu=1/(8 T_s)$, $\omega_{max}=2 \times 10^{-4}$, $\omega_{min}=\omega_{max}/2$. The crosses indicate the values of $<{E}_{x_1}>$ and $<{E}_{A}>$ obtained for $N=5,6,7,8,9,10,12,15$, and, $18$. \label{dim}}
\end{figure}

We tested our strategy for increasing network size $N$ up to $N=18$. As shown in Fig.  {\ref{dim}}, we found that our strategy was effective in identifying the network evolution for $5 \leq N \leq 18$. Note that for our choice of fully connected networks, the total number of unknowns to be  identified is equal to $N(N-1)$, e.g., it is equal to $20$ when $N=5$ and is equal to $306$ when $N=18$. We expect that, the task of identifying the evolution of the couplings becomes more difficult as the network dimension $N$ is increased. Consider for example a globally connected network of $N=18$ nodes. In such a case, each one of the nodes has to identify the time evolutions of $(N-1)=17$ different couplings to the remaining nodes in the network, from only one available received signal. Note that, by making the assumption of dealing with fully connected networks, and for which all the links are time-varying, we are considering a worst case scenario; in the case of our sensor application (see Sec. I), we expect that our strategy could benefit from a situation in which only a fraction of the network links might be changing as an object crosses the surveyed region. 

We also tried to estimate the time evolution of globally connected networks whose dimension $N$ is larger than $18$. For an $N=20$ node network, our strategy failed at identifying the couplings evolution; however, we repeated the same experiment with  $N=20$, but with the network evolving more slowly (i.e., $\omega_{max}=10^{-4}$), and in this case we were able to correctly identify the time evolving couplings. 

\subsection{Sensitivity to non-identicality of the network systems}

\begin{figure}[h]
\centerline{\psfig{figure=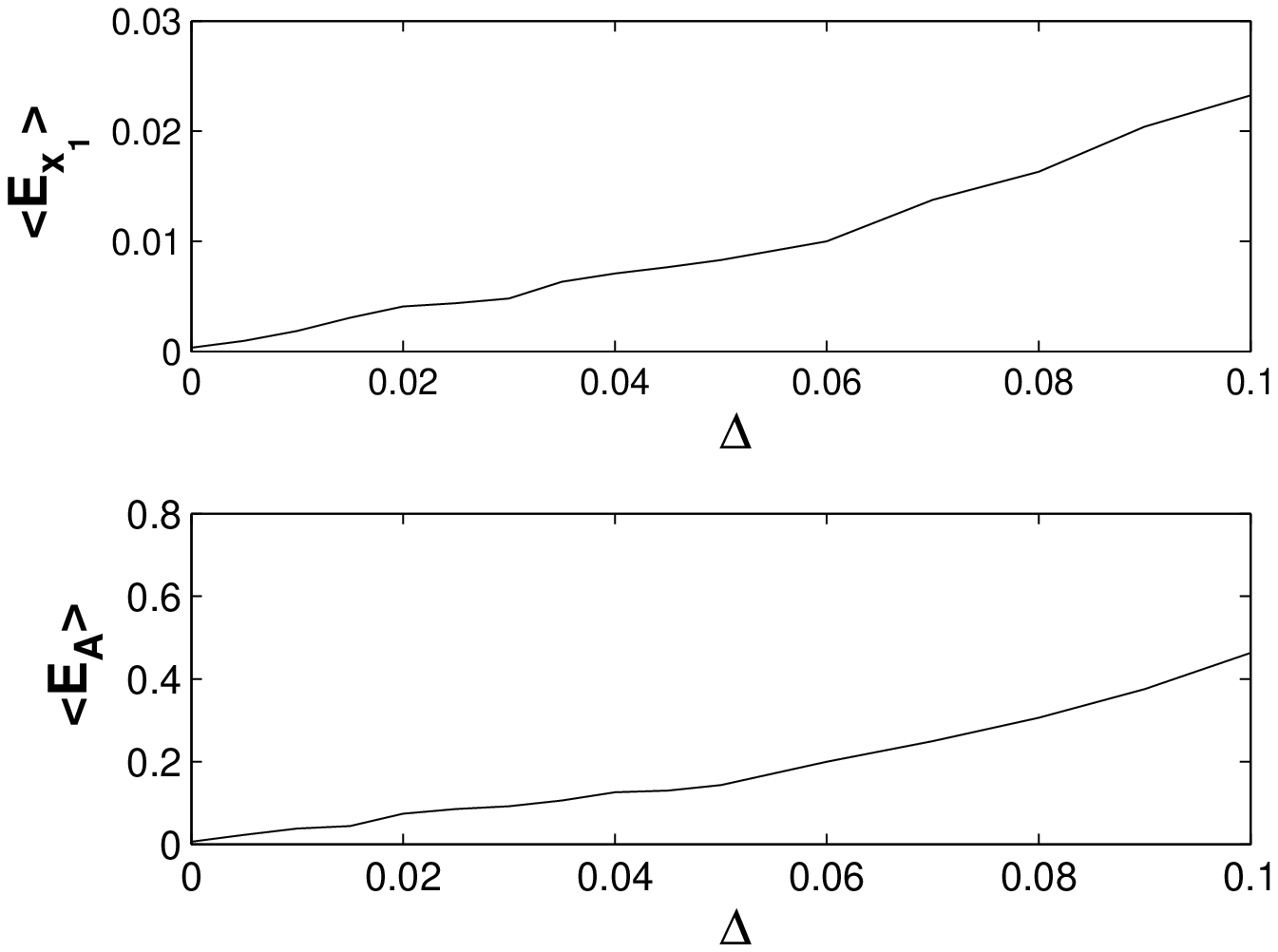,width=9cm,height=5cm}}
\caption{\small The time averaged errors $<{E}_{x_1}>$ and $<{E}_{A}>$ versus the non-identicality parameter $\Delta$, for a network of $N=5$ nodes,   $\alpha=0.2$, $\Gamma_1=1$, $\theta=0.6$, $\nu=1/(8 T_s)$, $\omega_{max}=2 \times 10^{-4}$, $\omega_{min}=\omega_{max}/2$. \label{Delta}}
\end{figure}

In practical situations (including our reference application of sensor networks) it is impossible to guarantee that all the network systems are precisely identical.
Therefore, we have also tested the robustness of our scheme to deviations of the individual systems from identicality. To this end, we replace $F(x_i)$ in Eq. (\ref{F}) by
\begin{equation}
F_i(x_i)=\left[\begin{array}{c}
    -x_{i2}-x_{i3} \\
    x_{i1} + 0.165 x_{i2} (1+ \Delta \delta_i)\\
    0.2+ (x_{i1}-10 )  x_{i3}
  \end{array} \right], \label{F}
\end{equation}
where for each node $i$ the parameter $\delta_i$ is chosen randomly with uniform density in the interval $[-1,1]$. We then repeat our original experiment. 
The parameter $\Delta$ characterizes the degree of non-identicality of the node dynamical systems. In Fig. \ref{Delta}, we show how the time averaged errors $<{E}_{x_1}>$ and $<{E}_{A}>$ depend on $\Delta$, for a network of $N=5$ nodes, $\alpha=0.2$. As can be seen, the more non-identical the systems are, the worse our strategy performs. However, for small values of $\Delta$,
for example, for $\Delta <0.03$, the synchronization and identification errors are respectively less than $0.5 \%$ and $10 \%$, i.e., $<{E}_{x_1}> \lesssim 5 \times 10^{-3}$ and $<{E}_{A}> \lesssim 0.1$, thus indicating that good results may still be obtained when the coupled systems deviate from being precisely identical.

\subsection{Sensitivity to errors in the values of $A_{iM}$}

\begin{figure}[h]
\centerline{\psfig{figure=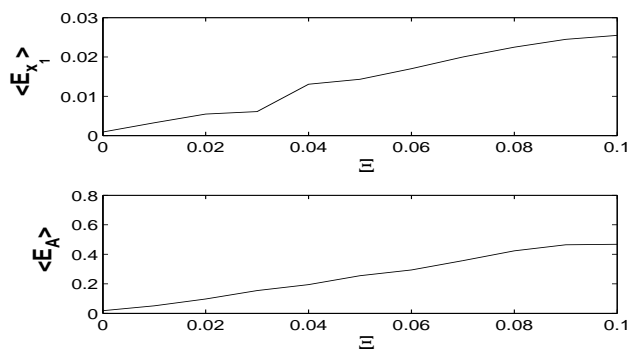,width=9cm,height=5cm}}
\caption{\small The time averaged errors $<{E}_{x_1}>$ and $<{E}_{A}>$ versus the mismatch parameter $\Xi$, for a network of $N=5$ nodes,   $\alpha=0.2$, $\Gamma_1=1$, $\theta=0.6$, $\nu=1/(8 T_s)$, $\omega_{max}=2 \times 10^{-4}$, $\omega_{min}=\omega_{max}/2$. \label{fcsi} }
\end{figure}

The results presented previously rely on the assumption that
the coupling with the maestro is exactly known at the network systems. Here we investigate a situation where the assumed known coupling with the maestro has a small error, and we study the effects of this error on our strategy. To this aim we replace Eq. (\ref{NP2}) by
\begin{equation}
\dot x_i(t) =  F(x_i(t)) + \Gamma  \{ r_i -[\frac{\alpha}{N-1} \sum_{j=1, j \neq i}^{N} \bar{A}_{ij} H (x_i(t-\tau_{ij})) + (1-\alpha) \tilde{A}_{iM} H(x_i(t-\tau_{iM}))]\}  ,  \quad i=1,2,...,N, \label{NP2bis},
\end{equation}
where $\tilde{A}_{iM}={A}_{iM}(1+ \Xi \delta_i)$, with $\delta_i$ is randomly chosen in $[-1,1]$, and the parameter $\Xi$ characterizes the degree to which the real coupling with the maestro $A_{iM}$ deviates from the coupling, $\tilde{A}_{iM}$, assumed at node $i$.
As can be seen from Fig. \ref{fcsi}, the larger the error, the worse our strategy performs. However, for small values of $\Xi$,
for example, for $\Xi <0.03$, the synchronization and identification errors are respectively less than $1 \%$ and $15 \%$, i.e., $<{E}_{x_1}> \lesssim   10^{-2}$ and $<{E}_{A}> \lesssim 0.15$, thus indicating that good results may still be obtained provided that the coupling with the maestro is affected by a not too large error.

We have also investigated the dependence of our strategy on the parameter $\alpha$ for the case in which small mismatches in the coupling with the maestro are present, i.e., $\Xi=0.02$. This is shown by the dotted line in Fig. \ref{alpha}. As can be seen, for such a case, our strategy works in a bounded interval of values of the parameter $\alpha$, i.e., $0.1 \lesssim \alpha \lesssim 0.45$, indicating that under these conditions, the best setup is realized when synchronization depends on both the influence of the maestro, as well as that of the orchestra.

\section{Conclusion}

In this paper we have dealt with the problem of identifying the time-evolutions of the couplings (assumed unknown) associated with the links of a network by making use of the phenomenon of chaos synchronization. We considered networks with time varying strengths of the couplings (due the effects of external unpredictable factors), as well as different coupling delays affecting the communication between the nodes.

 We proposed an adaptive strategy based on the concept of a `potential' that the network systems seek to minimize in order to identify the time evolutions of the network couplings. We considered fully connected networks of $N$ nodes and showed how our strategy could be successfully applied to identify the $N-1$ unknown couplings at each of the network nodes.
We emphasize that, we achieved this goal by using very little information exchanged at the network nodes: namely the coupling was realized in only one state variable, and the only available information at each node at time $t$ was an aggregate incoming signal from the other nodes, $r_i(t)$ (compare for instance our approach with others previously reported in the literature \cite{Identif1,Identif2}). Here we took advantage both of 
the phenomenon of synchronization of chaos and of 
the temporal signal diversity of the signal $r_i(t)$ due to the presence of \emph{different} time delays 
over the network links. 

We proposed two alternative versions of our strategy. In the first one, synchronization had to be achieved at each node (as well as the identification of the couplings over the network links), based solely on the information received from the other nodes in the network. 
For this case, we showed that we were able to correctly identify the evolution of the couplings in the case of small networks, i.e., a network of $N=3$ nodes, whose time evolution was sufficiently slow.

In order to extend and improve our results, we introduced a second version of our strategy, where we made use of an additional node, termed the maestro, having the function of maintaining network synchronization. We made the assumption that the coupling with the maestro was known at the other network nodes and that the maestro was not affected by the dynamics of the other nodes in the network. 
We showed that such an arrangement could lead to much better results than in the case where the maestro was absent. In particular, we were able to identify the time evolution of networks that were larger and more rapidly evolving than was possible in the other case. 
Other possible setups of our strategy are also possible, e.g., a selective choice of the nodes to be connected to the maestro (pinning control).

\section{Appendix}

In our experiment in Figs. \ref{F1}, \ref{F2}, and \ref{F3} (Version 1, $\theta=0.3$) we observed that we were able to synchronize the network on an approximate synchronous evolution obeying (\ref{is}), but we were not able to correctly identify the time evolutions of the $A_{ij}(t)$'s. In what follows, we give an explanation for this finding. 
Our explanation is that, when the $\tau_{ij}$ are not large enough, it is possible to obtain approximate (\emph{not} exact) synchronization for a range of values of the $\bar{A}_{ij} \neq {A}_{ij}$. Although this synchronization is not exact, it is still good enough to imply that the minimum of the potential $\Psi_i$ is so broad as to make our procedure ineffective. To illustrate this, assume the network systems evolve synchronously, i.e., $x_1(t)=x_2(t)=x_N(t)=x(t)$, and that the function $H(x(t))$ and $x(t)$ are infinitely differentiable.
Expanding $H(x(t-\tau_{ij}))$ in a Mac-Laurin series in the parameter $\tau_{ij}$, we have
\begin{equation}
H(x(t-\tau_{ij}))= \sum_{n=0}^{\infty} {(-\tau_{ij})}^{n} \frac{1}{n!} \frac{d^n H(x(t))}{dt^n}. \label{expans}
\end{equation}
If this expansion of the function $H(x(t-\tau))$ has a finite radius of convergence, then (\ref{expans}) is valid under the assumption that $\tau_{ij}$ is small enough.

Our strategy seeks to minimize the potential $\Psi_i$ at each node $i$, where $\Psi_i$ may be thought of as a sliding average of the mean squared synchronization error at node $i$, and the synchronization error is
\begin{equation}
Y_i(t)=  r_i(t) - \sum_{j \neq i}^{N} \bar{A}_{ij}(t) H( x_i(t-\tau_{ij})).
\end{equation}
Assuming $x_i(t)=x_j(t)$ for all $i$ and $j$ and making use of (\ref{expans}), we obtain
\begin{equation}
Y_i(t)= \sum_{n=0}^{\infty} \frac{(-1)^n}{n!} \frac{d^n H(x(t))}{dt^n}  \sum_{j=1, j \neq i}^{N}  [{A}_{ij}(t) - \bar{A}_{ij}(t)]  {\tau_{ij}}^{n}  \label{APP3}
\end{equation}
By construction, the $\bar{A}_{ij}(t)$ change slowly with time (on the time scale $\nu^{-1}$), as do the $A_{ij}(t)$ (on the time scale $T_n$). On the other hand, $x(t)$, and hence $\frac{d^n H(x(t))}{dt^n}$ change rapidly with time. Thus to make the sliding average of $Y_i(t)^2$ as small as possible, we need
\begin{equation}
\bar{A}_{ij}(t)=A_{ij}(t), \label{AA}
\end{equation}
giving zero for the sliding average. However, if $N$ is large and the $\tau_{ij}$ are not too big, we can achieve values of $\Psi_i$ that are very small, even if (\ref{AA}) is far from being satisfied. To see this we note that
%
if the delays $\tau_{ij}$ are too small, for some sufficiently large $n$, say $n^{*}$, the summation over $n$ may be truncated with only a small resulting error in $Y_i$. Assuming this to have been done, and setting each term in the sum over $n$ in (\ref{APP3}) equal to zero for each $n \leq n^*$, we obtain $n^{*}N$ equations for the $N(N-1)$ unknowns $\bar{A}_{ij}$.
If the number of unknowns is greater than $n^*$, then we can satisfy all the $n^*$ equations for a continuum of $\bar{A}_{ij}$ values not satisfying (\ref{AA}), and the potential $\Psi_i$ will then be near zero.

This work was supported by an MURI contract administered through ONR (contract N000014-07-0734).

\end{document}